\documentclass[a4paper,12pt,reqno,superscriptaddress,showkeys,nofootinbib]{revtex4}
	\usepackage[centertags]{amsmath}
	\usepackage{amsfonts}
	\usepackage{amssymb}
	\usepackage{amsthm}
	\usepackage{newlfont}
	\usepackage{stmaryrd}
	\usepackage{mathrsfs}
	\usepackage{mathtools}
	\usepackage{euscript}
	\usepackage{graphicx}
	\usepackage{enumerate}
	\usepackage{todonotes}
	\usepackage{comment}
	\usepackage{placeins}
	\usepackage{color}
	\usepackage{floatrow}
	\usepackage{caption}

	\usepackage{tikz}
	\usepackage{pgf}
	\usetikzlibrary{positioning,fit,calc}
	\usetikzlibrary{arrows,automata}
	\usepackage{wrapfig}
	\usepackage{subfigure}
	\usepackage{amscd}
	\usepackage{hyperref}
	
	\theoremstyle{plain}

	\theoremstyle{definition}

	\theoremstyle{remark}

	\newcommand{\opunit}{\text{1}\kern-0.22em\text{l}}

	\DeclareMathAlphabet{\mathpzc}{OT1}{pzc}{m}{it}

\begin{document}
		
\title{Death and resurrection of a current\\ by disorder, interaction or activity}

\author{Thibaut Demaerel and Christian Maes\\
	{\it Instituut voor Theoretische Fysica, KU Leuven}}
\begin{abstract}
	Because of disorder the current-field characteristic may show a first order phase transition as function of the field, at which the current jumps to zero when the driving exceeds a threshold. The discontinuity is caused by adding a finite correlation length in the disorder.  At the same time the current may resurrect when the field is modulated in time, also discontinuously: a little shaking enables the current to jump up.  Finally, in trapping models exclusion between particles postpones or even avoids the current from dying, while attraction may enhance it.  We present simple models that illustrate those dynamical phase transitions in detail, and that allow full mathematical control.
\end{abstract}

\maketitle

\maketitle

\section{Introduction}
Dynamical phase transitions in general refer to abrupt changes in macroscopic dynamical properties.  A relevant observable is the current in a nonequilibrium system, which depending on various external conditions and internal parameters may for example jump between different values, or be either strictly zero or nonzero. There is a large literature on the discussion of such transitions in terms of current fluctuations, especially starting from the nontrivial phase diagram for asymmetric exclusion processes; see \cite{krug} and e.g. the recent \cite{laz,kaf} and references therein.  An approach which starts from modifying the weight of path-space histories is much related and brings the subject also closer to the nature of glassy dynamics \cite{dpt} and kinetically constrained dynamics \cite{kin,chan}.   Obviously, to allow for nonanalytic behavior in the current characteristic (e.g. as function of the driving) we need to take some thermodynamic limit, but the dynamical transitions are visible already for sufficiently large systems as it should to be observable in real experimental set-ups. \\

In the following section we present random walk models where the current or speed shows a first order phase transition: strictly above a threshold in the driving field, the current tumbles to zero in a discontinuous way. In Section \ref{trapp} we show how adding activity via a time-dependence in the field (``shaking") may make the current nonzero in regimes where it vanishes otherwise.  Then, in Section \ref{inter} we consider the effect of particle interactions and specifically the influence of exclusion {\it versus} inclusion on the behavior of the current.  Clearly many-body effects do have an influence on localization, see e.g. \cite{gar,tra}.\\
In all cases, proofs can be made mathematically rigorous but in the present paper we mostly sketch the more heuristic arguments for the claims.\\

The general context has been considered before for a multitude of reasons.  Random walks or diffusions in random media is of course a famous subject covering a vast range of domains; see e.g. \cite{zei,snit,hug}, and we cannot even start giving a proper introduction or a more complete list of references. Another very related connection is to glassy behavior, ageing and/or anomalous diffusion in trapping models.  Again the literature is immense and we only recall some aspects in the beginning of Section \ref{trapp}. This paper emphasizes the aspect of dynamical phase transition in a number of simple models where the novelty is mostly in the (1) example of an overdamped diffusion in a random potential with a first--order transition as function of the field, (2) in the idea of resurrection of the current as function of the dynamical activity, here induced by (even a little) shaking, and (3) how all that may be modified depending on the nature of repulsive {\it versus} attractive interaction between the particles.  The models allow a complete mathematical treatment.

\section{The sudden death of a current}

One of the first theoretical physics examples where a particle current was seen to die beyond a treshold value for the external field is in \cite{Barma}.  That paper deals mainly with a two-dimensional set-up where the square lattice is diluted via independent bond removal.  When there is percolation of bonds a particle is driven through the (infinite) percolation cluster and it is seen numerically that the current (strictly) dies when pushing too hard. In this context, we agree to say that a function $f:(0,+\infty) \to \mathbb{R}$ ``dies" when $f$ is nonzero on an interval of the form $(0,x_d)$ with $x_d<\infty$ and it vanishes on $(x_d,+\infty)$.\\ 
The situation is of course reminiscent of the Lorentz gas with randomly placed obstacles, where, for the same physical reasons, a negative differential conductivity can be observed on finite graphs.  By now there is a large literature on such phenomena; see e.g. \cite{fre,slapik,ben,bae,kolk,Zia,sol}.\\

Still in \cite{Barma}, two different mechanisms are identified as responsible for the suppression of the current: first there is the possibility that the traversable ``backbones", i.e. the paths that remain open after the completion of the bond-removal process, display kinks. A particle will then have a hard time to move beyond the kink, since the external bias discourages this. Second, there are dead-ends attached to the backbones, i.e. paths with only one entrance. Particles may waste some time there before resuming the useful part of their journey. 

\subsection{The effect of disorder along the backbone} \label{death1}
To probe the effect of the kinks, Dhar and Bharma consider a one-dimensional and simpler version of the process. Here is a simple example in the same vein:\\

Let $w=(w_x, x\in \mathbb{Z})$ be independent $\{0,1\}$-random variables, where $w_x=1$ with probability  $\rho$, and $w_x=0$ with probability $1-\rho$, for fixed density $\rho \in [1/2, 1]$.  Given a realization of that field $w$, we introduce the driving parameter $p \in [1/2, 1]$ and we consider the discrete--time nearest--neighbor random walk $X_t$ on $\mathbb{Z}$ with transition probabilities
\begin{eqnarray}
\text{Prob}[X_{t+1}=x+1|X_t=x] &=& (2p-1) \,w_x + (1-p)\nonumber\\
\text{Prob}[X_{t+1}=x-1|X_t=x] &=& (1-2p) \,w_x + p
\end{eqnarray}
Since $\rho\geq 1/2$ there is a bias to the right when $p>1/2$.
That is a typical example of a  RWRE in dimension 1 as studied by Solomon \cite{sol} for which the speed 
\[
v(\rho,p) = \lim_t  \frac 1{t}\, X_t 
\]
exists and is explicit.  The result is that when $1/2<p < \rho$, there is the nonzero speed
\[
v(\rho,p)  = \frac{(2p-1)(\rho-p)}{ \rho (1-p) + p (1-\rho)  } >0     
\]
while, if $p\geq \rho$, then $v(\rho,p) =0$.  In other words, the current dies (continuously) at $p=\rho$.  There remains the linear response regime where $v(\rho,p) \simeq \varepsilon (2\rho-1)$ is increasing for small $\varepsilon =p -1/2>0$ and $\rho>1/2$, but then, for higher values of $p$, the current reaches a maximum and decreases to zero at $p=\rho$.\\
Note that the opposite phenomenon is also possible, as in depinning transitions \cite{tim1,tim2,tim3}, or as in the case of a yield stress where a solid starts to show plastic deformation and flows only beyond a certain stress.  In such and other similar cases, the corresponding diffusion constant remains zero then for small but finite driving, and the system only starts to ``move'' for large enough external field where also the Sutherland-Einstein relation between mobility and diffusion constant strictly fails. See also \cite{naim} for a discussion on anomalous diffusion due to disorder and/or interaction.\\
Observe also that {\it a priori} the disorder can work both on the bond or on the site variables.  Below, starting from equation \eqref{EOM}, we work with the bond version, where the ``force'' is random (entropic disorder).  In that case (bond disorder), there does need to be a correlation between going forward and backward; for site disorder it is the local escape rate which is variable (frenetic disorder).\\

There are also simple examples of overdamped diffusive systems with a current that goes to zero at some value of the external force.  Here however and as a new aspect to the theory and to the phenomena introduced above, we present a case, as in the title of the present section, where the current goes to zero with a jump (discontinuously).\\
Consider $x_t \in \mathbb{R}$ and the overdamped diffusion at times $t\geq 0$,
\begin{equation}\label{EOM}
\dot{x}_t = F(x_t) + \sqrt{2}\, \xi_t, \qquad \;\;x_0=0
\end{equation}
for standard white noise $\xi_t$. We put $k_BT=1$ with $T$ the temperature as energy reference, and the friction $\gamma=1$ also for convenience. The force $F= - V'$ and the potential $V$ is a random, continuous and piecewise smooth potential on $\mathbb{R}$ in such a way that its increments $V(x+\Delta x)-V(x)$ have a distribution that does not depend on $x$. For the sake of simplicity we continue here with one specific choice, while the mathematical treatment allows a much larger class of examples.\\

\begin{figure}[ht]
	\caption{Top picture: The random potential discussed in the text. Here $z_1=z_3=z_6=-1$ and $z_2=z_4=z_5=1$. Lower picture: the channel-interpretation corresponding to the potential above.}
	\centering
	\includegraphics[width=16cm]{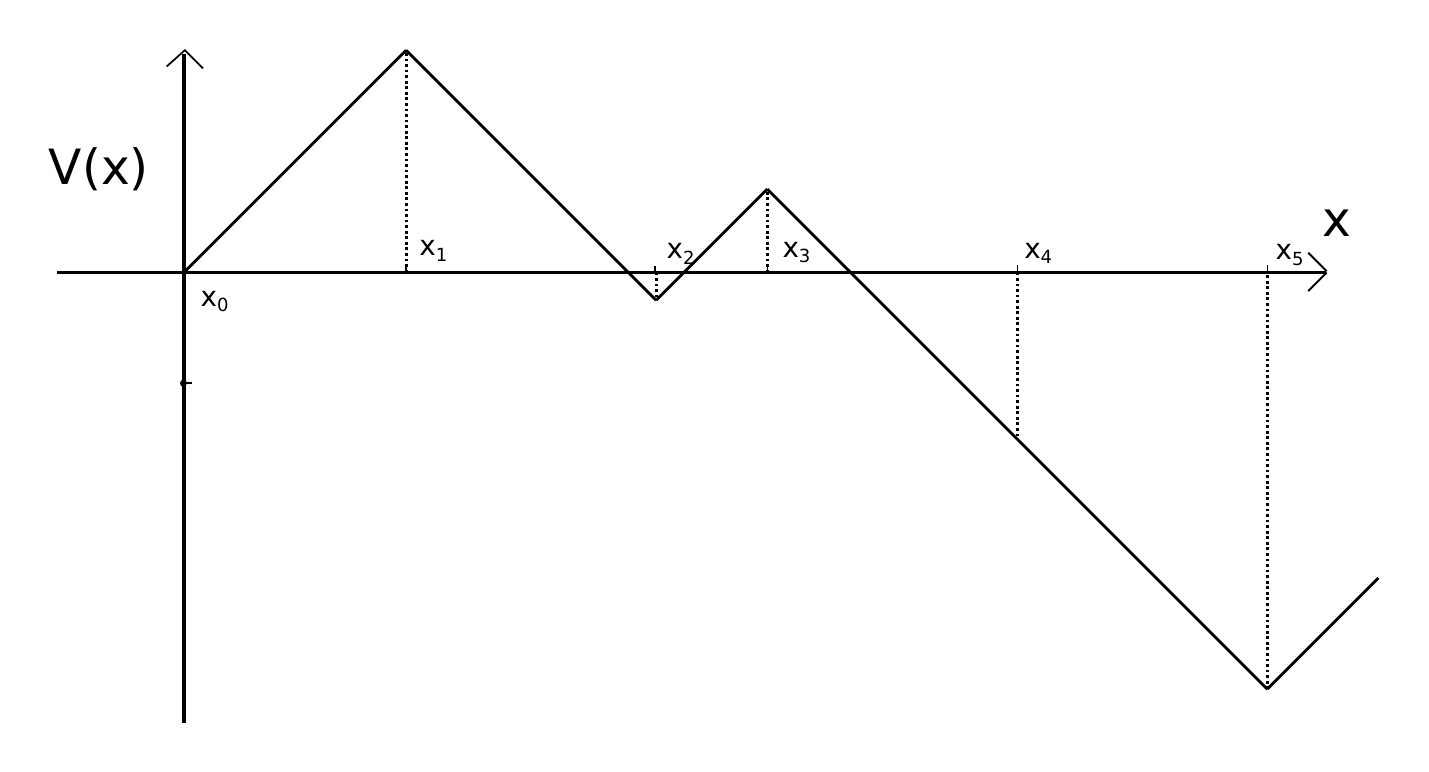}
	\includegraphics[width=16cm]{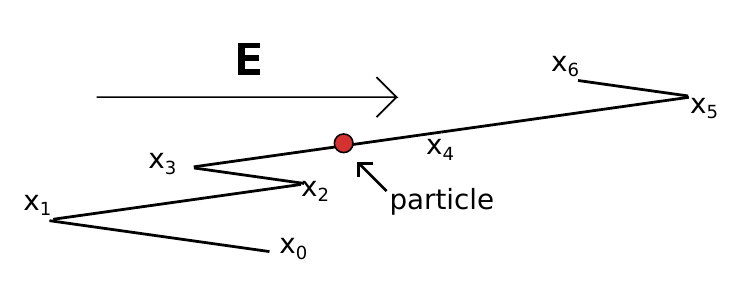}
	\label{dra}
\end{figure}

We take a parameter $w\in (0,1)$ to make a Bernoulli sequence $z_i = 1$ with probability $w$ and $z_i = -1$ with probability $1-w$ for $i\in\mathbb{Z}_0$.  When $z_i=+1$ we associate to it a length $\ell_i>0$ which is drawn with probability density
\begin{equation}
\label{rhoplus}
\rho_+(\ell)=\frac{1}{\ell_0\,{\cal L}(\nu\ell_0)}\frac{e^{-\nu \ell}}{1+(\ell/\ell_0)^2},\quad 
{\cal L}(s)=\int_0^{+\infty}\frac{e^{-s \,u}}{1+u^2}\,\text{d}  u
\end{equation}
with parameters $\nu,\ell_0>0$.  For $\ell_0\uparrow \infty$ the random length has an exponential distribution with average $\nu^{-1}$.  See Fig.~\ref{dra}.  The potential $V$ on $\mathbb{R}$ is obtained from associating to $i$ a slope $E>0$ and length $\ell_i$ when $z_i=1$, and a slope $-E$ and length $L$ (fixed) when $z_i=-1$. By fixing $V(0)=0$ we can construct the whole potential that way, and the force  $F=-V'$ is defined by joining the slopes. 
For large $w$ (close to one) there will be long intervals where the slope in $V$ is upward and the force points to the left.
More explicitly, we define $x_0=0$ and
\begin{eqnarray}
x_i &=& \sum_{j=1}^i [\ell_j\, \delta_{z_j,1} + L\,\delta_{z_j,-1}], \quad i=1,2,\dots\nonumber\\
x_i &=& -\sum_{j=i}^{-1} [\ell_j \,\delta_{z_j,1} + L\,\delta_{z_j,-1}], \quad i=-1,-2,\dots
\end{eqnarray}
so that the force equals
\begin{eqnarray}
F(x) &=& E \;\;\text { when }\,\;\; x\in (x_{i-1},x_i) \;\text{ for } \;\;z_i = 1, i=1,2\ldots\nonumber\\
F(x) &=& -E \;\;\text { when }\,\;\; x\in (x_{i-1},x_i) \;\text{ for } \;\;z_i = -1, i=1,2,\ldots
\end{eqnarray}
and similarly for $x<0$.\\
By choosing $L$ large enough we make sure that the potential tilts downward on average in the direction of increasing $x$.  The average force to the right is indeed
\[
\langle F\rangle = E\;\frac{(1-w)L - w \langle \ell\rangle }{(1-w)L + w\langle \ell\rangle }, \qquad \text{with}\;\;\; \langle \ell\rangle = -\ell_0\,\frac{\text{d}\log {\cal L}}{\text{d} s}(\nu\,\ell_0)
\]  

\begin{figure}[ht]
	\caption{The speed-field-characteristic for $w=0.005$, $0.05$, $0.1$, $0.3$ resp. and $L=\nu=\ell_0=1$. Computed via \eqref{speed}, \eqref{time'}, \eqref{list}}
	\centering
	\includegraphics[width=16cm]{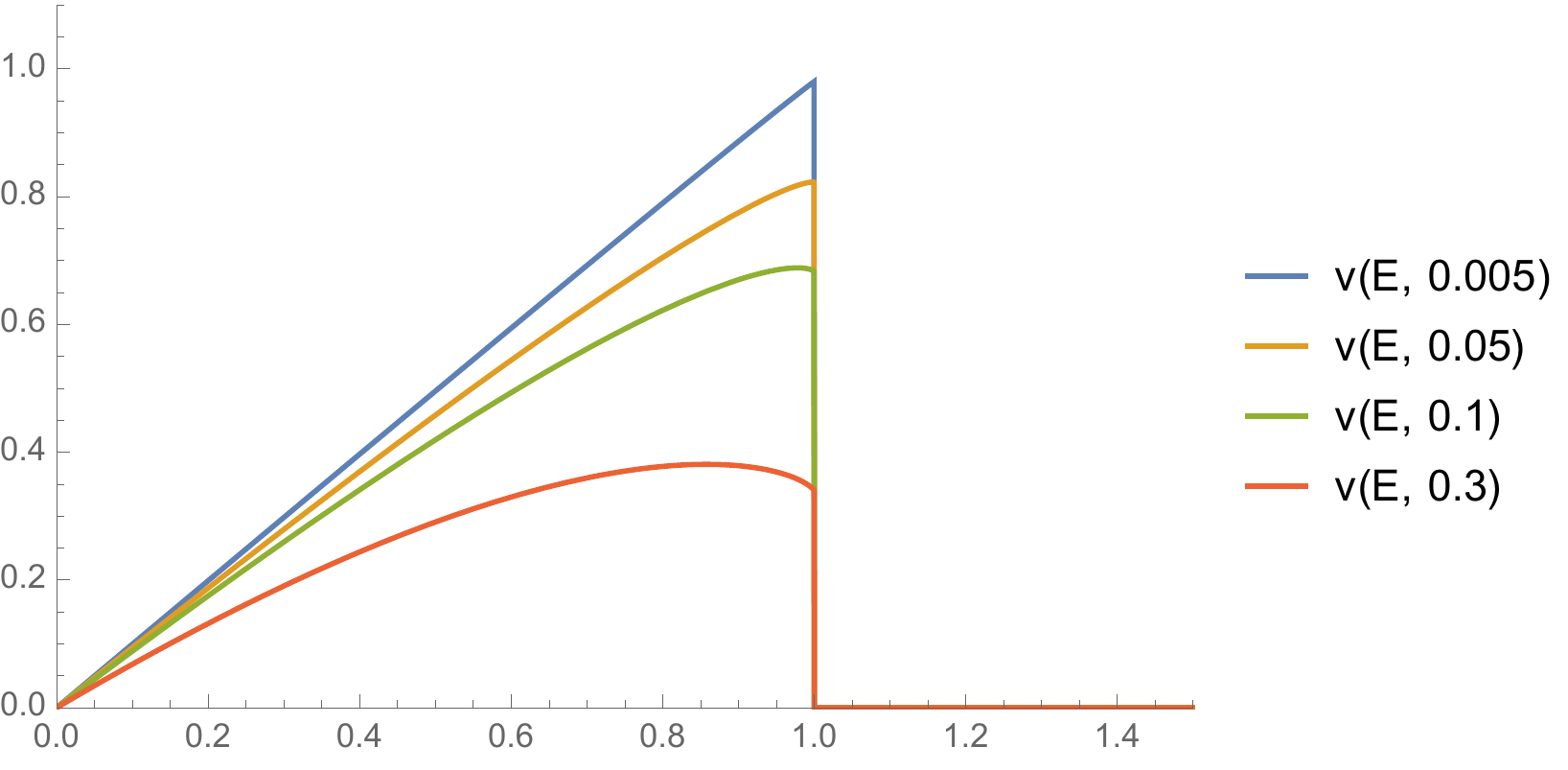}
	\label{cu}
\end{figure}

\noindent We have a net driving to the right for large enough $L$.  The interpretation where the randomness is associated to the geometry of a one-dimensional channel is displayed in the lower part of Fig.\ref{dra}. This clarifies also the link with the kinks considered in \cite{Barma}. The intervals $[x_{i-1},x_{i}]$ are either pointing to the right (in the direction of the external field $E>0$) when $z_i=-1$, or pointing to the left (against the external field) when $z_i=+1$. 
In that picture, the external field always points to the right with amplitude $E$ but by the geometry of the channel, particles need to move against the field on some (random) stretches of length $\ell_i$ during their journey.  In other words, the points $x_{i}$ become traps when $z_i=+1, z_{i+1}=-1$, and more so for large $E$. Trapping may of course also happen on a larger scale, in a combination or sequence of smaller traps as in Fig.~\ref{dra}. It is then not so surprising that the current (to the right) dies for large $E$ or for small $\nu$.
As we will show the current in fact goes to zero at $E = \nu$ with a jump, i.e., in a right-discontinuous way (see Fig. \ref{cu}).\\

The argument can be sketched as follows: If the expected length of the random intervals $[x_j,x_{j+1}]$ is bounded (as it is), then using the law of large numbers one can prove that the asymptotic speed 
\begin{equation}\label{speed}
v=\limsup_{ t \to \infty} \frac{x(t)}{t}\overset{a.s.}{=} \frac{\langle x_1-x_0\rangle}{ \langle \Delta t_0\rangle}=\frac{\langle x_1\rangle}{ \langle \Delta t_0\rangle}
\end{equation}
where $\langle \Delta t_0\rangle$ is the average time required for the particle to hit $x_1$ if it starts at $x_0=0$
It is therefore essential to compute the expected time (first passage time) $\langle \Delta t_j\rangle$ to cross an interval $[x_j,x_{j+1}]$.  Averaging (only) over the dynamics we have Dynkin's formula \cite{oks,sid}, 
\begin{equation}
\langle \Delta t_j\rangle = \int_{x_j}^{x_{j+1}} \text{d}  y\int_{-\infty}^{y} \text{d}  z \,e^{V(y)-V(z)}
\end{equation}
Note that the right-hand side has the dimension of length-squared = time, as one can see from the equation of motion \eqref{EOM} where the diffusion constant is taken dimensionless.\\
We rewrite this as
\begin{eqnarray*}
	&& \langle \Delta t_j\rangle = \int_{x_j}^{x_{j+1}} \text{d}  y\left\{e^{V(y)-V(x_j)}\underbrace{\int_{-\infty}^{x_j}\text{d}  z\,e^{V(x_j)-V(z)}}_{\kappa_j}+\int_{x_j}^y\text{d}  z\,e^{V(y)-V(z)}\right\} \\
	&& =\kappa_j \underbrace{\int_{x_j}^{x_{j+1}} \text{d}  y\,e^{V(y)-V(x_j)}}_{=:B_j}+\underbrace{\int_{x_j}^{x_{j+1}} \text{d}  y\int_{x_j}^y\text{d}  z\,e^{V(y)-V(z)}}_{=:C_j}
\end{eqnarray*}
Here $(B_j)_j$ and $(C_j)_j$ are iid, $(\kappa_j)_j$ is stationary, and $\kappa_j$ and $B_j$ are mutually independent.\\
Next we take the expectation over the disorder,
\[
\overline{\langle \Delta t_j\rangle}=\kappa B + C\]
To determine $\kappa $, note that
\begin{eqnarray*}
	&& \kappa_{j+1}=\int_{-\infty}^{x_{j+1}}\text{d}  z\, e^{V(x_{j+1})-V(z)} \\
	&& = \underbrace{e^{V(x_{j+1})-V(x_j)}}_{=:c_j}\int_{-\infty}^{x_j}\text{d}  z\, e^{V(x_j)-V(z)}+\underbrace{\int_{x_j}^{x_{j+1}} \text{d}  z \,e^{V(x_{j+1})-V(y)}}_{=:A_j} \\
	&& = c_j \kappa_j + A_j
\end{eqnarray*}
where $c_j$ and $A_j$ are iid, and $c_j$ and $\kappa_j$ are mutually independent.
Taking expectations on both sides yields $\kappa = c \kappa + A$. Hence (since $\kappa_j$ is a positive random variable),
\[\kappa = \begin{cases}
\frac{A}{1-c} & \text{ when } c<1\\
+ \infty & \text{ otherwise }
\end{cases}\]
We conclude that
\begin{equation}\label{time'}\overline{\langle \Delta t_j\rangle}=\begin{cases}
\frac{AB}{1-c}+C & \text{ when } c<1\\
+\infty & \text{ otherwise.}
\end{cases}\end{equation}

To see where the asymptotic speed is nonzero, we thus need to check where $c<1$, and to verify when $A, B, C$ are bounded:
\begin{equation}\label{list}\begin{cases}
& c(E)=(1-w) \,e^{-E L} + w\,\int_0^\infty\text{d}  \ell \rho_+(\ell)\,e^{E\ell}\\
& A= \frac{1-w}{E}e^{-E L}(e^{E L}-1) + \frac{w}{E}\int_0^\infty\text{d}  \ell\rho_+(\ell) e^{E \ell}(1-e^{-E \ell})\\
& B=A \\
& C=(1-w)\,(\frac{e^{-E L}-1}{E^2}+\frac{L}{E}) + \frac{w}{E}\int_0^\infty\text{d} \ell\rho_+(\ell)[\frac{1}{E}(e^{E \ell}-1)-L]
\end{cases}\end{equation}
We see that $c(E) <1$ for $E$ small enough. In fact,
\begin{equation}
c(E)=\begin{cases}
(1-w)\,e^{-E L}+\frac{w}{\ell_0{\cal L}(\nu\ell_0)}\int_0^\infty \frac{e^{(E-\nu)\ell}}{1+(\ell/\ell_0)^2}\,\text{d}  \ell& \text{ when }E\leq \nu \\
+\infty & \text{ otherwise}
\end{cases}
\end{equation}
For fixed $\nu$ and $\ell_0$ we can choose $w>0$ very small and $L$ sufficiently large so that $c(E) < 1$ on the interval $[0,\nu]$. Also $A$, $B$ and $C$ are, as a function of $E$, bounded on the interval $[0,\nu]$.  We conclude that the speed goes to zero at $E=\nu$ in a right-discontinuous way (see Fig. \ref{cu}).  It is also clear from the calculation (1) why it goes to zero at all, and (2) why it goes to zero with a jump.  Ad (1) there is little new except that there are few and simple diffusion models where the death of a current is shown in that generality.  Ad (2) there is something new here and the origin of the jump (and first-order transition) can be traced back to the finite exponential moment
\[
\langle e^{\nu \ell} \rangle =  \frac{\pi}{2{\cal L}(\nu\,\ell_0)} < \infty
\]
by the finiteness of $\ell_0$. That $\ell_0$ being finite means that the disorder deciding the length of the positive slopes in the potential $V$ is not entirely multiplicative: the (more microscopic) steps in the disorder are not independent and the large $\ell-$decay of the length of the ``bad'' domains follows a type of  Ornstein--Zernike law (polynomial correction to exponential decay).

\FloatBarrier
\subsection{The role of dead-ends for the diffusion of independent particles} \label{death2}
\begin{figure}[ht]
	\centering
	\includegraphics[width=14cm]{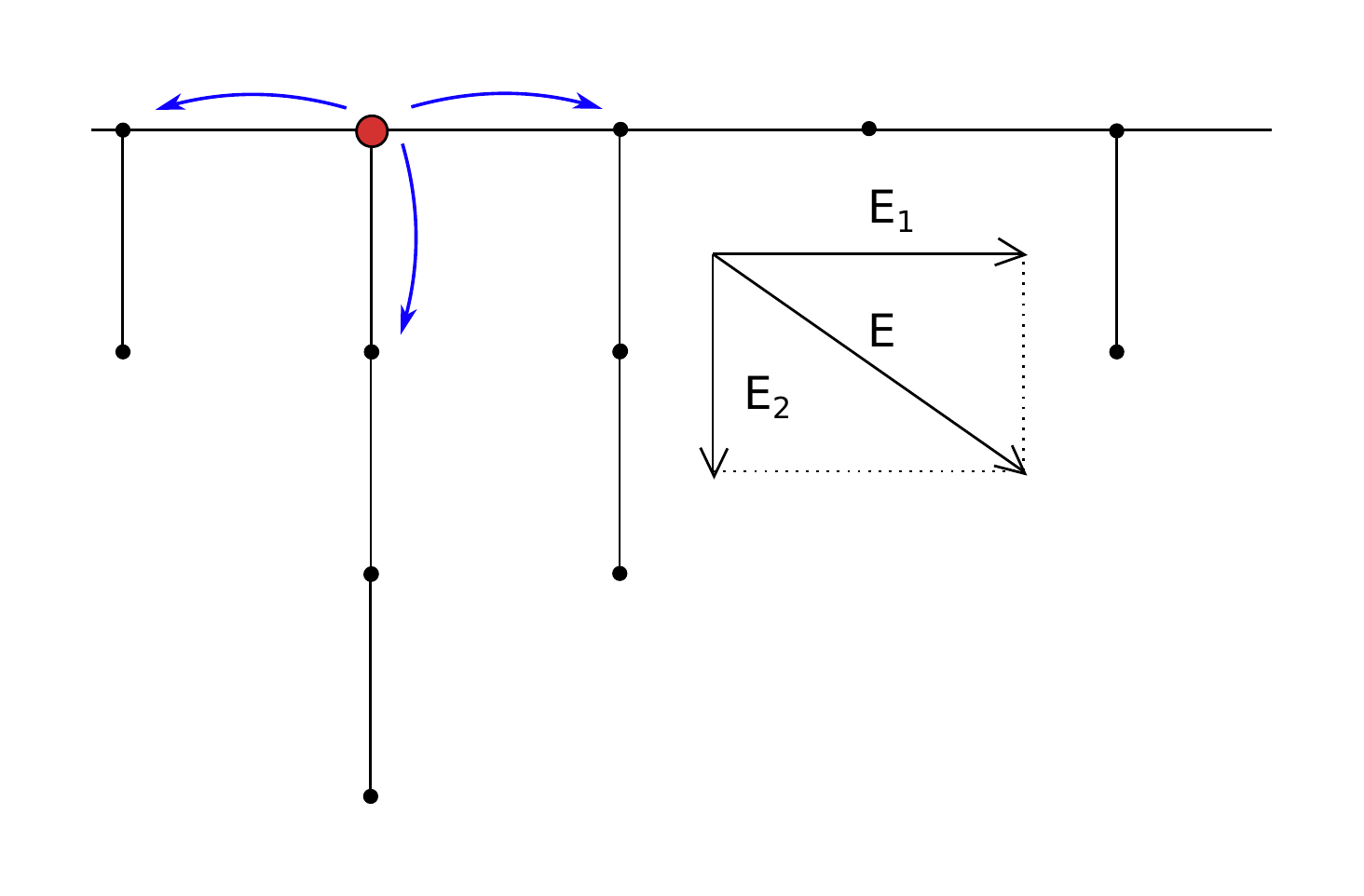}
	\caption{Walking on a random comb}
	\label{tr}
\end{figure}
In Section \ref{death1} we already outlined a scenario how kinks in the backbone may give rise to a first-order transition in the speed of the particle. In this section, we look at a continuous time-discrete space RW. The state space $S$ is arranged in a ``comb"-geometry, as a cartoon for a backbone with dead-ends. We refer to Fig.~\ref{tr}. More precisely, the sites are on $M:=\{(n,m)\in \mathbb{Z}^2\left.\right|0\leq m \leq L_n \}$ where the $(L_n)_n$ are positive integers, independent and identically distributed with finite expectation, $\mathbb{E}[L_n] <\infty$.\\
The inter-site jump rates are
\begin{itemize}
\item $e^{\pm E_1/2}$ for jumps from site $(n,0)$ to $(n\pm1,0)$.
\item $e^{ E_2/2}$ for jumps from site $(n,m)$ to $(n,m+1)$ (provided $m\leq L_n-1$ of course). 
\item $e^{-E_2/2}$ for jumps from site $(n,m)$ to $(n,m-1)$(provided $m> 0$ of course).
\item All other inter-site jumps are forbidden.
\end{itemize}
We ask what aspect in the distribution of the random variable $L_0$ could lead to a first-order phase transition in the speed
\begin{equation}
v:=\limsup_{t \to \infty}\frac{n(t)}{t}
\end{equation}
Basically one proceeds by proving the three successive equalities
\begin{equation} \label{eq}
v\overset{a.s.}{=}\frac{1}{\langle\Delta t_0\rangle}=\tanh( {\cal E}_1/2)\frac{1}{\langle\tilde{\Delta t_0}\rangle}=2\sinh( {\cal E}_1/2)\frac{e^{  {\cal E}_2}-1}{ e^{ {\cal E}_2}\langle e^{ L_0 {\cal E}_2}\rangle-1}
\end{equation}
where $\Delta t_0$ is the time required for the walker to reach the site $(1,0)$ for the first time when it starts at the site $(0,0)$ and the average $\langle... \rangle$ is both over the random comb-geometry and the jump dynamics. Likewise, $\langle \tilde{\Delta t_0} \rangle$ is the average time required for the walker to reach either $(-1,0)$ or $(1,0)$ given that it started at $(0,0)$. The first equality of \eqref{eq} is a consequence of the law of large numbers (for stationary processes). The second equality follows from the following argument:
\begin{enumerate}
	\item The average time for the walker to reach either $(-1,0)$ or $(1,0)$ given that it started at $(0,0)$ is $\langle\tilde{\Delta t_0}\rangle$ as said before
	\item The probability for the walker to first jump to $(\pm1,0)$ is resp. $\frac{e^{{\cal E}_1/2}}{e^{{\cal E}_1/2}+e^{-{\cal E}_1/2}}$ and $\frac{e^{-{\cal E}_1/2}}{e^{{\cal E}_1/2}+e^{-{\cal E}_1/2}}$.
	\item In the former case, we had $\tilde{\Delta t_0}=\Delta t_0$. In the latter case, the particle jumped to $(-1,0)$ after a time $\tilde{\Delta t_0}$. From there, it has to wait (on average) a time $\langle\Delta t_{-1}\rangle=\langle\Delta t_0\rangle$ to jump back to $(0,0)$ and then an addition time $\langle\Delta t_0\rangle$ to finally jump to $(1,0)$. Therefore
	\[\langle\Delta t_0\rangle=\langle\tilde{\Delta t_0}\rangle+\frac{e^{-{\cal E}_1/2}}{e^{{\cal E}_1/2}+e^{-{\cal E}_1/2}}2\langle\Delta t_0\rangle,\]
	or
	\[\langle\Delta t_0\rangle=\frac{\langle\tilde{\Delta t_0}\rangle}{\tanh( {\cal E}_1)}.\]
\end{enumerate}
The third equality in \eqref{eq} is shown through a straightforward exercise in the theory of average escape times in a Markov jump process with finite state space.\\\\
Inspecting \eqref{eq}, we see again that the speed is zero either when ${\cal E}_1$ is zero or when $\langle e^{ L_0 {\cal E}_2}\rangle$ (which is a function ${\cal E}_2$) is zero. Very much similar as in Section \ref{death1} we find that when $L_0$ is distributed in a non-multiplicative way, e.g. with
\begin{equation}\label{nonmult}\text{Prob}\left[L_0=\ell\right]=\frac{1}{{\cal L}(\nu,L_0)}\frac{e^{-\nu L}}{1+(L/L_0)^2},\end{equation}
then the current goes to zero in a discontinuous way at ${\cal E}_2=\nu$. Note that a breaking of multiplicativity such as in \eqref{nonmult} is quite likely when considering the length-distribution of dead-ends in percolation clusters: simply the conditioning on a dead-end being a dead-end rather than a backbone does the job. 
\section{Resurrection of a current}\label{trapp}

Given the importance of dynamical activity in nonequilibrium physics \cite{diss}, we may wonder if shaking may make the current to resurrect. We proceed wth additional features related to the phenomenon that we discussed already in \cite{condmat}.\\  
At the same time we enter here the realm of trapping models which since many years (cf \cite{bou,mal}) have been central in questions of glassy behavior, ageing behavior or anomalous transport.  The trapping will here be realized by a crystalline comb-geometry.\\

\begin{figure}[ht]
	\centering
	\includegraphics[width=14cm]{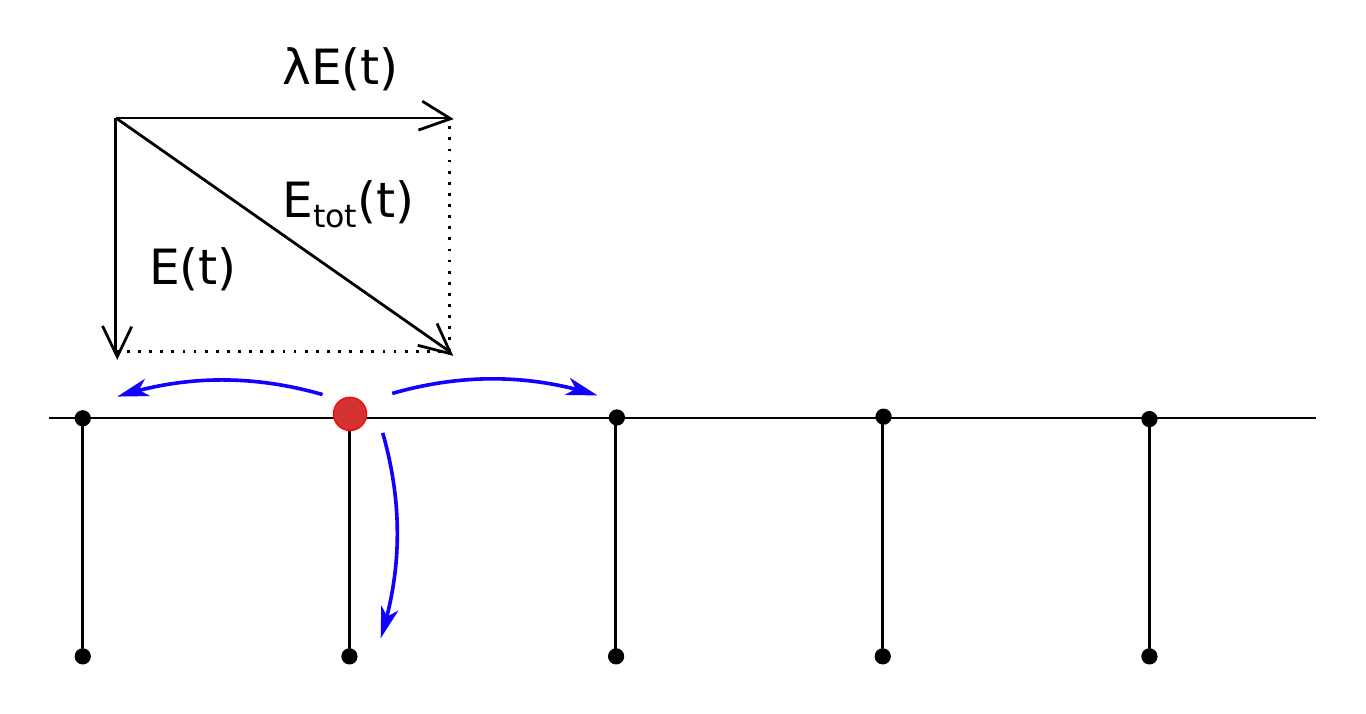}
	\caption{The active $a$-states are represented by the black dots on the backbone of the comb, while the sleeping $s$-states are represented by the black dots on the dents. The arrows on the upper left illustrate the applied force: its magnitude and direction may change in time, but its angle remains fixed. That is: the ratio of its horizontal and vertical components is $\lambda$ at all times.}
	\label{trap'}
\end{figure}
We consider a continuous  time random walk on the comb-like configuration space\\ $\mathbb{Z}_n\times \left\{a,s\right\}$ where $a$ stands for ``active'' and $s$ for ``sleeping" (a trapped state) and $\mathbb{Z}_n$ is the ring with $n$ sites; see Fig.~\ref{trap'} for a portion of that ring. The dynamics is a time-dependent Markov process.  The transitions are $(j,a)\rightarrow (j,s)$ at rate $q(t)$, $(j,s)\rightarrow (j,a)$ at rate $p(t)$.  Furthermore, to get a current we choose $\lambda>0$ and let $(j,a)\rightarrow (j+1,a)$ at rate $q^\lambda(t)$, $(j+1,a)\rightarrow (j,a)$ at rate $p^\lambda(t)$
for all $1\leq j \leq n$ (periodic).   The fact that the vertical and horizontal transition rates are coupled must be interpreted via the presence of an external field for which the angle with the vertical is quantified with $\lambda$.
We take $p(t)q(t) = 1 = p^{\lambda}(t)q^{\lambda}(t)$ to have a constant activity parameter.\\
With $\rho_a(t)$ the total probability (at time $t$) to be in an active state and $\rho_s(t)= 1-\rho_a(t)$ the probability to be sleeping, 
\begin{equation} \label{Master}
\begin{cases}
& \dot{\rho}_a(t)=p(t)\rho_s(t)-q(t)\rho_a(t)= p(t)(1-\rho_a(t))-q(t)\rho_a(t)\\
& \dot{\rho}_s(t)=q(t)\rho_a(t)-p(t)\rho_s(t)= q(t)(1-\rho_s(t))-p(t)\rho_s(t)
\end{cases}
\end{equation}
The instantaneous macroscopic current over the ring $J(t)$ is
\begin{equation}
J(t)=\rho_a(t)[q^{\lambda}(t)-p^{\lambda}(t)]
\end{equation}

The time-dependence is chosen such that  $p(t)=q^{-1}(t)$ is periodic with period $\tau=t_1+t_2$,
\[p(t)=\begin{cases}
p_1 & \text{ when } t\in [0,\,t_1) \\
p_2 & \text{ when } t\in [t_1,\,t_1+t_2)
\end{cases}\]
Writing the escape rates and jump probabilities as 
\begin{eqnarray*}
	&& \Lambda_1=p_1+q_1, \quad \Lambda_2=p_2+q_2 \\
	&& \rho_1 = \frac{p_1}{\Lambda_1},\;\; \;\;\qquad \rho_2 = \frac{p_2}{\Lambda_2}
\end{eqnarray*}
one easily checks that $\rho_a$ solving \eqref{Master} converges for large $t$ to the periodic function
\begin{equation}
\rho_a(t)=\begin{cases}
\frac{(\rho_2-\rho_1)(1-e^{-\Lambda_2t_2})}{1-e^{-\Lambda_1t_1-\Lambda_2t_2}}e^{-\Lambda_1 t}+\rho_1 & \text{ when }t\in [0,t_1]\mod t_1+t_2\\
\frac{(\rho_1-\rho_2)(1-e^{-\Lambda_1t_1})}{1-e^{-\Lambda_1t_1-\Lambda_2t_2}}e^{-\Lambda_2 (t-t_1)}+\rho_2 & \text{ when }t\in [t_1,t_1+t_2]\mod t_1+t_2
\end{cases}
\end{equation}

\begin{figure}[t]
	\centering
	\includegraphics[width=14cm]{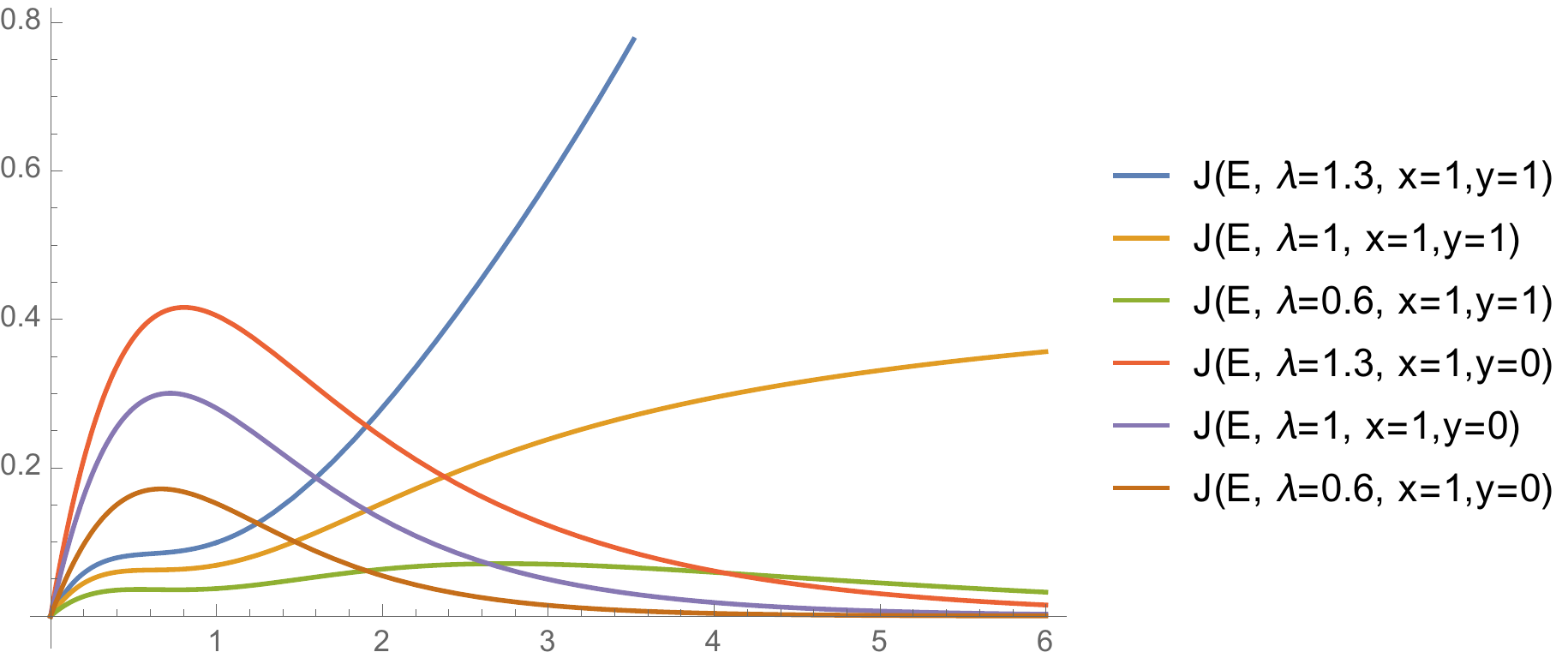}

	\caption{The shaken and unshaken currents ${\cal J}(E,x=1,y=1)$ resp. ${\cal J}(E,x=1,y=0)$
	as a function of $E$ for $\lambda=1.3,\lambda=1$ and $\lambda=0.6$ (the ordering in the legend agrees with the top-down order of the graphs at $E=6$).}
	\label{sha}
\end{figure}

In the steady state the average current ${\cal J}=\frac{1}{\tau}\int_0^\tau J(t)$ is given by ($x:=\Lambda_1t_1$, $y:=\Lambda_2t_2$)
\begin{eqnarray}
&& {\cal J}=\frac{1}{t_1+t_2}\left\{\int_0^{t_1} \rho_a(t')[q_1^{\lambda}-p_1^{\lambda}]\text{d} t'+\int_{t_1}^{t_1+t_2} \rho_a(t')[q_2^{\lambda}-p_2^{\lambda}]\text{d}  t' \right\} \\
&& =\frac{1}{(t_1+t_2)(1-e^{-(x+y)})}\left\{\frac{(\rho_2-\rho_1)(1-e^{-x})(1-e^{-y})+\rho_1x(1-e^{-(x+y)})}{\Lambda_1}[q_1^{\lambda}-p_1^{\lambda}]\right. \nonumber \\
&&  \qquad\qquad \left.+ \frac{(\rho_1-\rho_2)(1-e^{-x})(1-e^{-y})+\rho_2y(1-e^{-(x+y)})}{\Lambda_2}[q_2^{\lambda}-p_2^{\lambda}] \right\}
\end{eqnarray}
When we put $q_1=p_1^{-1}=e^{E}$ and $p_2=q_2=1$, we have $\Lambda_1=2\cosh E$, $\Lambda_2=2$, $\rho_1=\frac{e^{-E}}{2\cosh E}$ and 
$\rho_2=\frac{1}{2}$.
The expression for the average current ${\cal J} = {\cal J}(E,x,y)$ reduces to
\begin{equation} \label{current}
{\cal J}=\frac{\sinh(\lambda E)}{(x+2y\cosh E )(1-e^{-(x+y)})\cosh E}  \left\{(\cosh E -e^{-E})(1-e^{-x})(1-e^{-y})+e^{-E}x(1-e^{-(x+y)})\right\}  
\end{equation}

Putting $y=0$ in this result brings us to the result of the time-independent system where ${\cal J}(E,x,0)= e^{-E}\,\sinh(\lambda E)/\cosh E$, or
\[
{\cal J}(E,x,0) \simeq e^{E (\lambda-2)},\qquad E \uparrow \infty
\]
We see the familiar behavior of a current going to zero for large $E$ provided $0<\lambda < 2$.  The physical reason is that as $E$ increases the walker also goes to sleep more often.\\
However, if we keep $x,y>0$ fixed in \eqref{current}, one finds
\[
{\cal J}(E,x,y) \simeq e^{E (\lambda-1)}\,\frac{(1-e^{-x})(1-e^{-y})}{2y(1-e^{-(x+y)})},\qquad E \uparrow \infty
\]
so shaking can indeed ``resurrect'' the current for $\lambda\geq 1$; see Fig.~\ref{sha}.  For $\lambda \in (1,2)$ the current is exponentially smaller in $E$ compared to the case when activity is added in terms of shaking.  Of course the limit of $E\uparrow \infty$ cannot be exchanged with that of $y\downarrow 0$, and even for very small shaking (small $y$) the effect of resurrecting the current is immediately there.  Methods of enhancing flow or of restoring current are obviously of much practical importance.  Here and in \cite{condmat} we propose that a little shaking can do a lot; see also \cite{cat} in the case of granular suspensions.\\

Note that in the above example specific choices were made for the rates. More specifically: one can make other choices regarding the amount of traffic between the different states and how that depends on the external field. For virtually all choices one comes to the conclusion that some amount of shaking can significantly boost the current and overall activity. \\
Secondly, the resurrection (as the vanishing) of the current here plays at large $E$; mathematically in the limit $E \uparrow \infty$ only.  That is however the result of the uniform boundedness of the depth of the traps, i.e. of the dents in  Fig.~\ref{trap'}.\\  In the next section we concentrate on the influence of interaction but we also make the comb random.  In other words we will make the sleeping state more dynamical and resolve its microscopics.

\FloatBarrier
\section{Adding interaction}\label{inter}
So far we have studied the influence of disorder and activity on the current characteristic. Here we show how the nature of the interaction can also dramatically and differently influence the current.

\subsection{Exclusion enhancing the current} \label{hooks}

\subsubsection{Uniformly bounded traps}
\begin{figure}[ht]
	\label{trap}
	\centering
	\includegraphics[width=12cm]{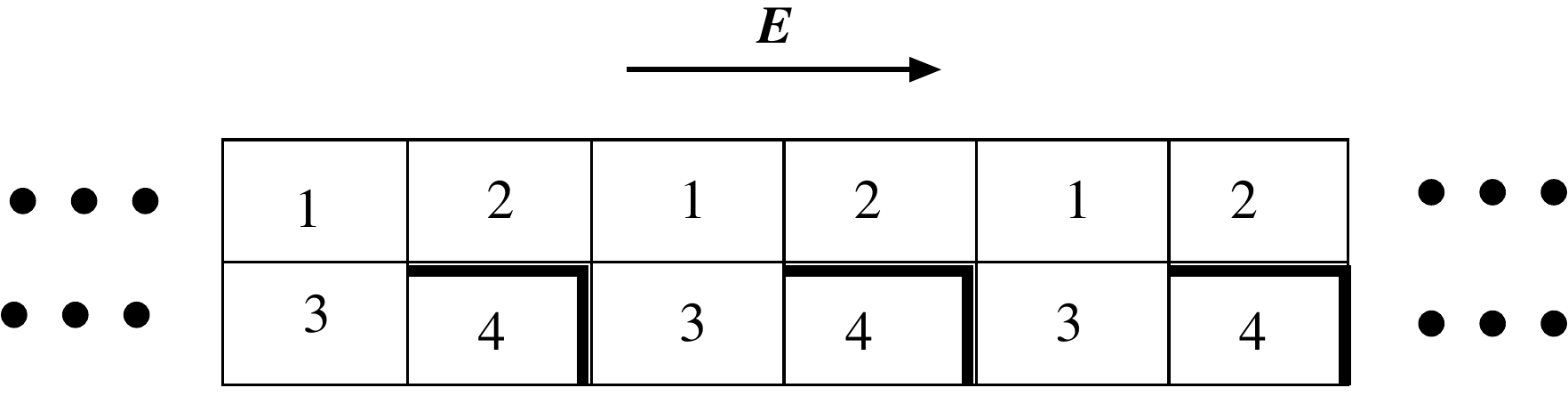}
	\caption{A comb like in Fig.~\ref{trap'} where the dents are bent inward. Figure courtesy by \cite{bae}.}
	\label{hook}
\end{figure}

Before we make the comb random we consider the set-up of \cite{Zia,bae}, but with exclusion. Consider therefore a system of $N$ mutually excluding particles in Fig.~\ref{hook}.  A cell has address $c = (\ell,n)$ where $1\leq \ell \leq L$ and $1\leq n \leq 4$. 

\begin{figure}[h]
	\centering
	\includegraphics[width=14cm]{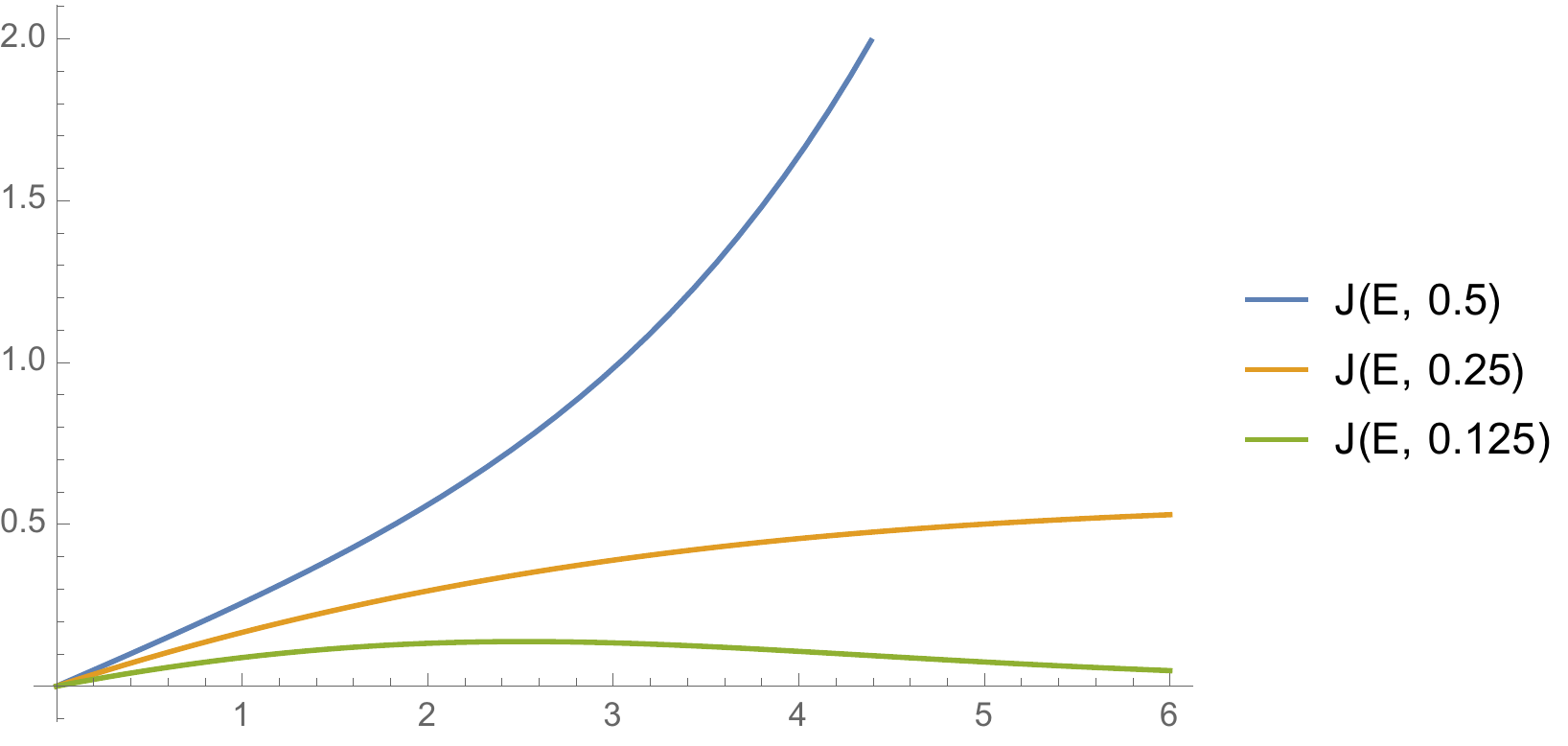}
	\caption{Plot of the current \eqref{curr} for the values $\rho=0.5$, $0.25$ and $0.125$ respectively.}
	\label{4cell}\end{figure}

The particles have labels (are distinguishable) and the state of the system is therefore $x = (x_1,\ldots,x_N)$ with each $x_i$ being some cell $c$.  For the current that will not matter.  The dynamics is by exchanging the occupation between neighboring cells.  Horizontal exchanges (if permitted) have rates $\exp \pm  E/2$; vertical transitions (if permitted) have rate 1.  The permission depends on there being a ``wall'' or not between adjacent cells; see the thick lines in Fig.~\ref{hook}.  When there would be no exclusion and the particles are independent, the current would become vanishingly small for large  $E$.\\

The stationary distribution is given by 
\begin{equation}
\text{Prob}(x)= \frac{1}{{\cal Z}}\exp\left(-\sum_{j=1}^N \phi(x_j)\right)
\end{equation}
where the potential $\phi(c)$ in a cell $c$ is defined by
\begin{equation}
\phi(c)=\phi((\ell,n))=\begin{cases}
0& \text{ when }n\in \left\{1,2,3\right\} \\
-E & \text{ otherwise.}
\end{cases}
\end{equation}
That stationary measure is a (Fermi-Dirac) product measure in the open finite system or in the thermodynamic limit.  Suppose indeed that particles are introduced both at $(1,1)$ and $(L,2)$ at rates $\frac{1}{1+\theta} e^{ E/2}$ and $\frac{1}{1+\theta} e^{- E/2}$ respectively, at least when those cells are not occupied by a particle already. Likewise, for the exit of particles at $(1,1)$ and at $(L,2)$  we take the rate $\frac{1}{1+\theta}$.  Then the stationary distribution is the grand-canonical one,
\begin{equation} \label{stat}
\text{Prob}(x) = \frac{1}{\Xi}\prod_{\text{cell }c}[\theta^{-1} e^{- \phi(c)}]^{\text{d} lta_{c \in x}}
\end{equation}
The correct normalization is 
\begin{equation}
\Xi = \prod_{\text{cell }c}[1+\theta^{-1}e^{- \phi(c)}]
\end{equation}
so that we can rewrite \eqref{stat} in the familiar Fermi-Dirac form:
\begin{equation}
\text{Prob}(x) = \prod_{\text{cell }c}\frac{[1+\theta e^{ \phi(c)}]^{1-\text{d} lta_{c \in x}}}{1+\theta e^{ \phi(c)}}.
\end{equation}
In the thermodynamic limit, a similar product distribution is stationary. In that setting, one adjusts the chemical potential $\theta$ to fit that measure to the desired average particle density.\\

In the 4-cell system, the $n=1,2,3$-cells are occupied by a particle with probability $\frac{1}{1+\theta}$. The cell $n=4$ is occupied with probability $\frac{1}{1+\theta e^{- E}}$.
Since the measure is a product-measure, the expected total number $4\rho$ of particles in one macroscopic unit is the sum of the expected number of particles in each sub-cell, i.e.,
\begin{equation}
4\rho = \frac{3}{1+\theta}+\frac{1}{1+\theta e^{- E}}
\end{equation}
If we now adjust $E$ while the density $\rho$ is to remain constant, $\theta$ becomes a function of $E$: 
\begin{equation}
\theta(E)=\frac{[e^{ E}+3-4\rho(e^{ E}+1)]+\sqrt{[e^{ E}+3-4\rho(e^{ E}+1)]^2+4(4-4\rho)4\rho e^{ E}}}{8\rho}
\end{equation}
When $\rho<1/4$, one has $\theta(E)\sim \frac{1-4\rho}{4\rho}e^{ E}$ when $E \to +\infty$. When $\rho=1/4$, it is $\theta(E)\sim \sqrt{3}\,e^{ E/2}$. Finally, when $1<4\rho<4$, $\theta(E) \to \frac{4-4\rho}{4\rho-1}$ when $E \to +\infty$. For the current $J$, one has
\begin{eqnarray}
&& J(E) =\text{Prob}(\text{cell }n=1 \text{ occupied})\text{Prob}(\text{cell }n=2 \text{ unoccupied})\,e^{ E/2} \nonumber\\
&& - \text{Prob}(\text{cell }n=2 \text{ occupied})\text{Prob}(\text{cell }n=1 \text{ unoccupied})\,e^{- E/2} \nonumber\\
&&=\frac{2\theta(E)}{(1+\theta(E))^2}\sinh( E/2) \label{curr}
\end{eqnarray}
Therefore,
\begin{equation}\label{tra}
\lim_{E \to \infty} J(E)=\begin{cases}
0& \text{ when }\rho<1/4, \\
\sqrt{1/3}& \text{ when }\rho=1/4, \\
+\infty & \text{ when } 1/4 <\rho<1
\end{cases}
\end{equation}

In other words, when the density becomes sufficiently elevated, the exclusion prevents the current from dying.
Note however that the current also depends on the choice of rates and would change if the ``time-scale'' would also depend on $E$.  For example, the dynamical activity or traffic in a cell also changes with $E$.  The sharp transition in \eqref{tra} is on one specific time-scale, but the general feature of exclusion promoting current seems more universal.
\FloatBarrier
\subsubsection{ASEP on random combs} \label{ASEP}

\begin{figure}[ht]
	\centering
	\includegraphics[width=14cm]{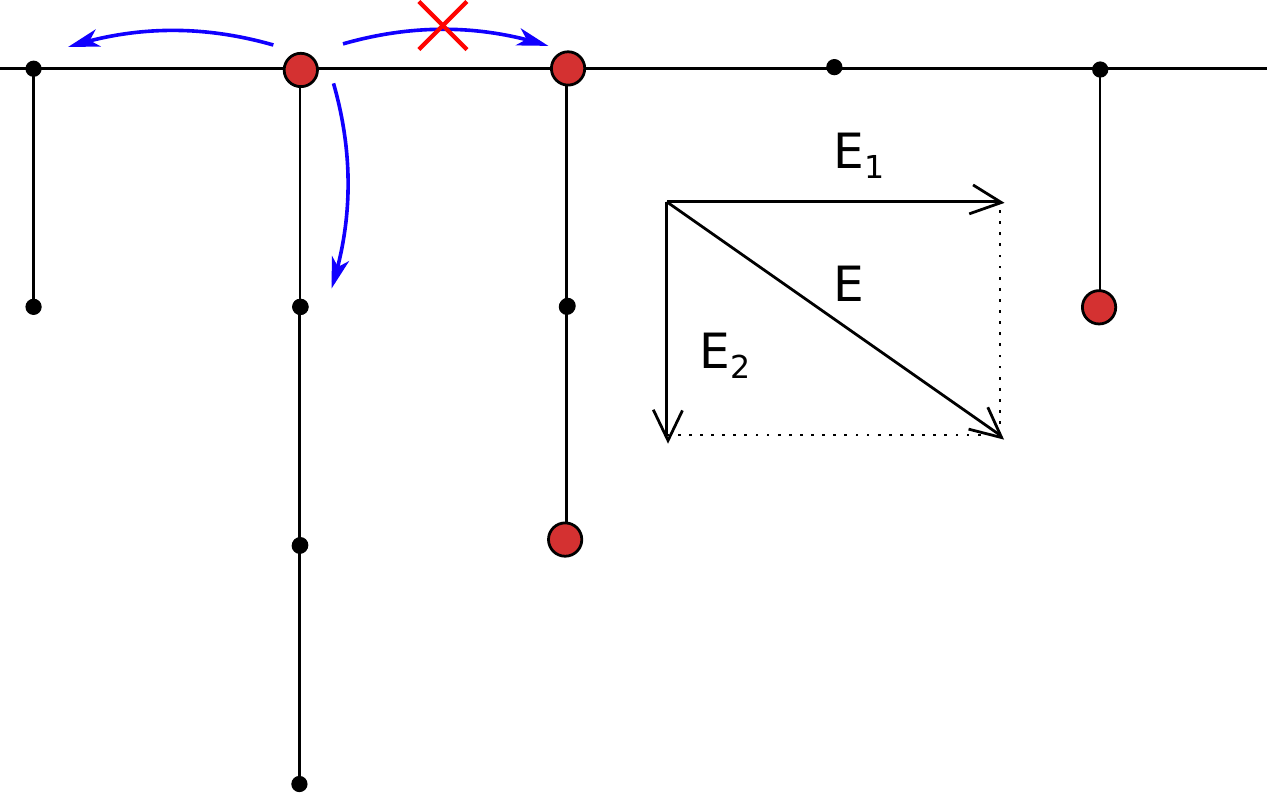}
	\caption{The asymmetric simple exclusion process on a random comb. }
	\label{combi}\end{figure}

We continue with the study of an asymmetric exclusion process on a comb whose dents have a random (integer) length. We refer to Fig.~\ref{combi}. More precisely, the sites are on $M:=\{(n,m)\in \mathbb{Z}^2\left.\right|0\leq m \leq L_n \}$ where the $(L_n)_n$ are positive integers, independent and identically distributed with finite expectation, $\mathbb{E}[L_n] <\infty$.\\
The jump rates imposed on individual particles are identical to those considered for the random walk in section \ref{death2}, but of course we now impose that such jumps are annulled in case the destination-site of the jump is already occupied by a particle. \\  

Consider the single site occupation densities $\nu_{\theta,E_2}^s$ defined as
\begin{eqnarray*}
	&&\nu_{\theta,E_2}^s[\text{$s=(n,m)$ is occupied by a particle}]=\frac{1}{1+\theta e^{- m E_2}} \\
	&&\nu_{\theta,E_2}^s[\text{$s=(n,m)$ is not occupied by a particle}]=\frac{\theta e^{- m E_2}}{1+\theta e^{- m E_2}},
\end{eqnarray*}
One can show that the product probability measure $\mu_{\theta,E_2}=\otimes_{s \in M}\nu_{\theta,E_2}^s$ is invariant under the dynamics.\\

The fugacity $\theta$ determines the overall particle density
\begin{eqnarray}
&& \rho(\theta,E_2)=\frac{\langle\text{ number of particles on the sites }\cup_{m=0}^{L_0}(0,m)\rangle}{\mathbb{E}[L_0]}\\
&& =\frac{1}{\mathbb{E}[L_0]} \sum_{j=0}^{\infty} \text{Prob}[L_0=j]\sum_{m=0}^{j}\frac{1}{1+\theta e^{- m E_2}}
\end{eqnarray}
The translational (horizontal) current is 
\begin{eqnarray}\label{tran}
&& J_{\theta}(E_1) = \text{Prob}[\text{site $(0,0)$ occupied and site $(1,0)$ unoccupied}]e^{ E_1/2} \nonumber\\
&& \qquad \;\;\; - \;\;\text{Prob}[\text{site $(0,0)$ unoccupied and site $(1,0)$ occupied}]e^{- E_1/2}\nonumber\\
&&=\frac{\theta}{(1+\theta)^2} (e^{ E_1/2}-e^{- E_1/2}) 
\end{eqnarray}
The current vanishes in the limits $\theta \to 0$ and $\theta \to +\infty$. The first limit corresponds to the density approaching 1 (mathematically as a consequence of the monotone convergence theorem); the other limit corresponds to the density going to zero (by the dominated convergence theorem).  To understand that $J_{\theta}(E_1) > 0$ for all $E_1>0$, we need to see that for nontrivial densities $0<\rho<1$ there is a nonzero and finite fugacity.  To arrive at that we notice that $\theta \mapsto \rho(\theta,E_2)$ is differentiable 
with respect to $\theta$ to any order.  The inverse function theorem implies that the strictly decreasing function $\rho(\theta,E_2)$ can be inverted with respect to $\theta$ to obtain a strictly decreasing function $\theta(\rho,E_2)$. Hence, if $0<\rho<1$, then $0<\theta(\rho,E_2)<\infty$ and therefore $J_{\theta(\rho,E_2)}(E_1)>0$ provided $E_1>0$.
We conclude that the exclusion between the particles assures that an external field {\it always} gives rise to a nonzero current. When there is no such exclusion (independent particles) and the dents have for example an exponential distribution, there is a threshold in the external field above which the current strictly vanishes.
\FloatBarrier
\subsection{Attractive zero-range process on random combs}


Suppose now that particles are allowed to jump to already occupied sites. We take a zero range process with constant rates.  That means that the dynamics of the previous section described in 2(a)--(b) still applies for precisely one particle on every occupied site; additional particles, if present, are inactive until said particle has left the site.\\
In that case the product probability measure $\mu_{\theta,E_2}=\otimes_{s \in M}\nu_{\theta,E_2}^s$ is invariant under the dynamics, where $\nu_{\theta,E_2}^s$ is defined as
\begin{equation*}
\nu_{\theta,E_2}^s[\text{$s=(n,m)$ is occupied by $k\in \mathbb{N}$ particles}]=\left(\theta e^{ m E_2}\right)^k\left[1-\theta e^{ m E_2}\right]
\end{equation*}
Note however that that is ill-defined unless $\theta\,e^{ mE_2}<1$ for all $m$. Since $m$ can take values in a random range, depending on the length of the dent we must require that the random variable $L_j \sim L_0$ has a finite support. So consider the above described asymmetric zero range process on a random comb where the dents are uniformly bounded by length $L_\text{max}$: for every fixed $E_2$ we have that $\mu_{\theta,E_2}$ is well-defined for all $\theta < e^{- L_\text{max} E_2}$. 

The translation (horizontal) current is
\begin{eqnarray*}J_{\theta}(E_1):=
	&&  \mathbb{E}_{\theta,E_2}[\text{Number of particles at site }(0,0)]e^{ E_1/2} \\
	&& -\mathbb{E}_{\theta,E_2}[\text{Number of particles at site }(1,0)]e^{- E_1/2} \\
	=&& \sum_{j=0}^{\infty}\left(j\theta^j(1-\theta)\right)(e^{ E_1/2}-e^{- E_1/2}) \\
	=&&\frac{\theta}{1-\theta}(e^{ E_1/2}-e^{- E_1/2}) < \frac{1}{e^{ L_\text{max} E_2}-1}(e^{ E_1/2}-e^{- E_1/2}) 
\end{eqnarray*}
We see that the current converges to zero if we let $L_\text{max} \to \infty$: the current gets vanishingly small as the support of the length distribution becomes infinite, for all fields $E_1$ and allowed densities. 
That is of course in sharp contrast with the exclusion process of the previous section where the current \eqref{tra} or \eqref{tran} will keep away from zero, uniformly in any cut-off $L_{\text{max}}$.

\section{Conclusion}

We have presented mathematically fully treatable models of phase transitions in the current.  For independent particles we gave two models where the current vanishes discontinuously beyond a certain field strength, either due to disorder along the backbone or due to trapping in dead-ends.  That threshold can be arbitrarily low when we add attraction as in the zero range process of the previous section, but for exclusion processes at high density there may not be a threshold at all.  We have also demonstrated how an arbitrary weak shaking in terms of a time-modulation can make the current nonzero in regimes where it would otherwise vanish.


\begin{thebibliography}{20}
	
	
	\bibitem{krug}
	J.~Krug, Boundary-induced phase transitions in driven diffusive systems. Phys. Rev. Lett. {\bf 67}, 1882 (1991).
	
	\bibitem{laz}
	A.~Lazarescu, Generic Dynamical Phase Transition in One-Dimensional Bulk-Driven Lattice Gases with Exclusion. J. Phys. A: Math. Theor. {\bf 50}, 254004 (2017).
	
	
	\bibitem{kaf}
	Y.~Baek, Y.~Kafri and V.~Lecomte, Dynamical phase transitions in the current distribution of driven diffusive channels. J. Phys. A: Math. Theor. {\bf 1}, 105001 (2018).
	
	
	\bibitem{dpt}
	J.P.~Garrahan, R.L.~Jack, V.~Lecomte, E.~Pitard, K.~van Duijvendijk, and F.~van Wijland,
	First-order dynamical phase transition in models of glasses: an approach based on ensembles
	of histories, J. Phys. A: Math. Gen. {\bf 42}, 075007 (2009).
	
	\bibitem{kin}
	J.P.~Garrahan, P.~Sollich, C.~Toninelli, Kinetically Constrained Models, arXiv:1009.6113.  In: L.~Berthier, G.~Biroli, J-P.~Bouchaud, L.~Cipelletti and W.~van Saarloos, editors, pages 341-369, Oxford University Press, 2011. 
		
	\bibitem{chan}
	R.~Jack, J.P.~Garrahan, D.~Chandler, Space-time thermodynamics and subsystem observables
	in kinetically constrained models of glassy materials. J. Chem. Phys. {\bf 125}, 184509 (2006).
			
	\bibitem{gar}
	B.~Everest, I.~Lesanovsky, J.P.~Garrahan and E.~Levi, Role of interactions in a dissipative many-body localized system. Phys. Rev. B {\bf 95}, 024310 (2017).
	
	\bibitem{tra}
	R.~Ramaswamy and M.~Barma, Transport in random networks in a field: interacting particles.
	J. Phys. A: Math. Gen. {\bf 20} (1987).	
	
	\bibitem{zei}
	O.~Zeitouni, Random walks in random environments.  Proceedings of the ICM, Beijing 2002, vol. 3, 117--130 (2003).
	
	\bibitem{snit}
	A.-S.~Sznitman, {\it Brownian Motion, Obstacles and Random Media}. Springer-Verlag --- Berlin, 1998.
	
	\bibitem{hug}
	B.D.~Hughes, {\it Random Walks and Randon Environments}, Volume 2: Random Environments.
	Oxford University Press, Oxford, UK (1995).
	
	\bibitem{Barma} 
	M.~Barma and D.~Dhar, Directed diffusion in a percolation network. J. Phys. C {\bf 16}, 1451 (1983).
	
	\bibitem{fre}
	S.~Leitmann and T.~Franosch, Nonlinear response in the driven lattice Lorentz gas. Phys. Rev. Lett. {\bf 111}, 190603 (2013).
	
	\bibitem{slapik}
	A.~Slapik, J.~Luczka and J.~Spiechowicz, Negative mobility of a Brownian particle: Strong damping regime. Commun. Nonlinear Sci. Numer. Simulat. {\bf 5}, 316--325 (2018).
	
	\bibitem{ben}
	O.~B\'enichou, P.~Illien, G.~Oshanin, A.~Sarracino and R.~Voituriez,
	Microscopic theory for negative differential mobility in crowded environments.
	Phys. Rev. Lett. {\bf 113}, 268002 (2014).
	
	\bibitem{bae}
	P.~Baerts, U.~Basu, C.~Maes and S.~Safaverdi, The frenetic origin of negative differential response. Physical Review E {\bf 88}, 052109 (2013).
	
	\bibitem{kolk}
	U.~Basu and C.~Maes, Nonequilibrium Response and Frenesy. J. Phys.: Conf. Ser. {\bf 638}, 012001 (2015).
	
	\bibitem{Zia}
	R.K.P.~Zia,  E.~L.~Pr{\ae}stgaard, and O.G.~Mouritsen, Getting more from pushing less: Negative specific heat and conductivity in nonequilibrium steady states. Am. J. Phys. {\bf 70}, 384 (2002). 
	
	\bibitem{sol}
	F.~Solomon, Random walks in a random environment. Ann. Prob. {\bf 3}, 1--31 (1975).
	
	\bibitem{tim1}
	A.~Larkin, Sov. Phys. JETP {\bf 31}, 784 (1970).
	
	\bibitem{tim2}
	H.~Leschhorn et L.-H.~Tang, Avalanches and correlations in driven interface
	depinning, Phys. Rev. E {\bf 49}, 1238--1245 (1994).
	
	\bibitem{tim3}
	T.~Thiery, Analytical Methods and Field Theory for Disordered Systems. Ph.D. Thesis at the Laboratoire de Physique Th\`eorique
	de l’Ecole Normale Sup\`erieure, 2016.
	
	\bibitem{naim}
	E.~Ben-Naim and P.L.~Krapivsky, Strong Mobility in Weakly Disordered Systems. Phys. Rev. Lett. {\bf 102}, 190602 (2009).
	
	\bibitem{oks}
	B.K.~Oksendal, {\it Stochastic Differential Equations: An Introduction with Applications }. Berlin: Springer, 2003.
	
	\bibitem{sid}
	S.~Redner, {\it A Guide to First-Passage Processes}. Cambridge University Press, 2001.
	
	\bibitem{diss}
	C.~Maes, {\it Non-Dissipative Effects in Nonequilibrium Systems}. SpringerBriefs in Complexity,  2018.
	
	\bibitem{condmat}
	T.~Demaerel and C.~Maes, Activity induced first order transition for the current in a disordered medium.  Condensed Matter Physics {\bf 0}, 33002 (2017). 
	
	\bibitem{bou}
	J.-P.~Bouchaud, Weak ergodicity breaking and aging in disordered systems.
	J. Phys. I (France) {\bf 2} , 1705–1713 (1992).
	
	\bibitem{mal}
	M.~Henkel and M.~Pleimling, {\it Non-Equilibrium Phase Transitions
		Volume 2: Ageing and Dynamical Scaling Far from Equilibrium}.  Springer, 2010.
	
	\bibitem{cat}
	C.~Ness, R.~Mari, M.~E.~Cates,
	Shaken and stirred: Random organization reduces viscosity and
	dissipation in granular suspensions.  Sci. Adv. {\bf 4} (3), eaar3296 (2018).
	
	
\end{thebibliography}
\end{document}